\documentstyle[twocolumn,aps,psfig,epsfig,floats]{revtex}
\begin{document}
\draft
\twocolumn[\hsize\textwidth\columnwidth\hsize\csname
@twocolumnfalse\endcsname

\title{Human Sexual Contact Network as a Bipartite Graph }
\author{G\"{u}ler Erg\"{u}n}
\address{Department of Mathematical Sciences, Brunel University,\\
Uxbridge, Middlesex UB8 3PH, U.K.}
\address{Electronic address: Guler.Ergun@brunel.ac.uk}
\maketitle \thispagestyle{empty}

\begin{abstract}
A simple model to encapsulate the essential growth properties of
\emph{the web of human sexual contacts} is presented. In the model
only heterosexual connection is considered and represented by a
random growing bipartite graph where both male-female contact
networks grow simultaneously. The time evolution of the model is
analysed by a rate equation approach leading to confirm that male
and female sexual contact distributions decay as power laws with
exponents depending on influx and charisma of the sexes.
\end{abstract}

 ] \narrowtext
%\vskip2pc]
%\begin{multicols}{2}
\pagebreak

%************************************************************
\section{Introduction}
\label{sec:intro}

Yet another human interaction network! \emph{The web of human
sexual contacts} is introduced in \cite{liljeros}, where the
authors make extensive analysis of 1996 Swedish survey of sexual
behaviour and show that the cumulative distributions of number of
sexual partners for males and females have power-law forms with
exponents $\alpha_{f}\approx 2.54$ for females and
$\alpha_{m}\approx 2.31$ for males, during a single year and
$\alpha_{f_{tot}}\approx 2.1$, $\alpha_{m_{tot}}\approx 1.6$ for
entire life time.

Human sexual behaviour is a subject area \cite{buss} in itself and
beyond the scope of this work. Here we  will concentrate on
capturing the essential details needed to form such a social
network for which we will make use of the developments from the
previous network models
\cite{barabasi,exactsol,topology,scaling,organization,review,review2}
and general consensus about human sexual behaviour.

The model introduced in \cite{barabasi} has identified two basic
principles that seem to govern the growth dynamics of human
interaction networks, namely preferential attachment and
continuous growth. This differs from the traditional complex
network consideration \cite{erdos} of connecting together a fixed
number of network elements (nodes) which results in Poisson
distributions of the connectivity.

The usual approach in most complex network models
\cite{review,review2,geoff,ergun} that have human attributes is to
employ the preferential attachment process; an incoming node is
more likely to connect to a node with higher connections. This
process yields power-law distributed connectivity in the network.
A similar process takes place in a sociophysics model
\cite{bonabeau}, where probability of winning a fight depends on
the number of victories, consequence of which is a self-organised
hierarchy in the system. Again a well studied biological evolution
model \cite{bak-sneppen}, in which the power-law behaviour is due
to increasing fitness of the species.

Contrary to previous human interaction network models, the model
we consider here has two different types of nodes and the linking
process is discriminative as well as preferential. In the
proceeding section, the model is built by connecting new nodes to
nodes with higher connections and analysed by a rate equation
approach. The results and the discussions are presented in the
final section.

%************************************************************
\section{The model}
\label{sec:growth}

The growth process of the sexual contact network described by a
random growing bipartite graph (illustrated in Fig.~\ref{fig1})
considers the following:
\begin{itemize}
 \item[(i)]With probability $q$ a new (no previous connection)
 female is connected to an existing male, which contributes to the number of
 females with contact degree $l=1$ in the network and increases the
 degree $k$ of the connecting male.
 \item[(ii)] With probability $p$ a new male is connected to
 an existing female, which has the analogous effect as in (i).
 \item[(iii)]With probability $r$ a link is created between an existing
 male and an existing female, which increases the degrees of both sexes.
\end{itemize}
In addition to above rules, initially there must be at least one
connected male-female pair and only heterosexual contact is
allowed among the sexes. Also the process $r$ only counts once
between the same pair; that is once a link between two individuals
is created it remains and further contact between them does not
increase their contact degrees.
\begin{figure}[h!]
\centerline{\epsfig{file=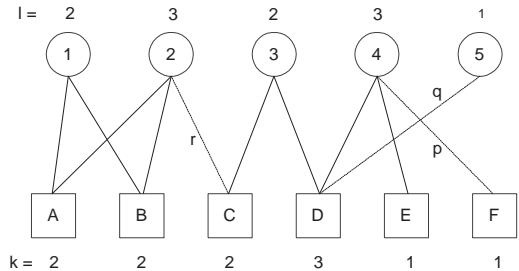,width=8.5cm}}
\caption{Schematic illustration of the growth processes of
male-female sexual contact network. Male nodes (squares) labelled
A to F and female nodes (circles) labelled 1 to 5. The numbers
above the nodes are indicating the node degree $l$ for females and
$k$ for males, i.e. female node $1$ has degree $2$ and male node
$D$ has degree $3$.}\label{fig1}
\end{figure}

We define male sexual degree distribution $A_{k}$ as the number of
males with $k$ female connections. The rate at which a male
attracts females is proportional to his number of connections and
it grows linearly with increasing $k$; in other words the
connection probability of a male increases as he becomes more
experienced. And similarly does the female sexual degree
distribution $B_{l}$ with attraction rate proportional to $l$.
However, it seems necessary to consider some characteristics
unique to both sexes so that we may underpin the social
expectations about their sexual contacts. Here we introduce
$\lambda$ and $\mu$ as the \emph{charisma} parameters of males and
females respectively. These parameters should not be confused with
\emph{charm} or \emph{beauty}, which could be used as
attractiveness of individuals. Here we are considering only the
general attractiveness of the sexes, which are shaped by social
expectations.

The charisma parameters are added into the rate of connections
given $k+\lambda$ for males and $l+\mu$ for females. Since we need
$k+\lambda>0$ and $l+\mu>0$ for the network to grow, $\lambda$ and
$\mu$ are considered to be positive integers and small compared to
the contact degrees of $k$ and $l$. Also, the general consensus is
that males tend to exaggerate their number of sexual partners.
This desire puts females in demand and increases female charisma.
Conversely, the social expectation of females restricts them to
report less and partly because of this they are more selective
about their sexual partners. Using this information we argue that
the rate at which females attract sexual partners should be
greater than that for males and therefore $\mu> \lambda$.

In this model we are considering a large system where the random
fluctuations in the linking process can be neglected when compared
with the collective behaviour of the individuals. Therefore it is
appropriate to use mean-field approximation. Then we can write the
time evolution of the average number of males with $k$ contacts
$A_{k}(t)$ as
\begin{eqnarray} \label{m-evolve}
\frac{d
A_{k}(t)}{dt}&=&\frac{q+r}{M_{a}}\left[(k-1+\lambda)A_{k-1}(t)-
(k+\lambda)A_{k}(t)\right]\nonumber \\&&+p\delta_{k1}
\end{eqnarray}
and similarly for females with $l$ contacts,
\begin{eqnarray}
\label{f-evolve} \frac{d
B_{l}(t)}{dt}&=&\frac{p+r}{M_{b}}\left[(l-1+\mu)B_{l-1}(t)-
(l+\mu)B_{l}(t)\right]\nonumber \\&&+q\delta_{l1}.
\end{eqnarray}
Since the structure of Eq.~(\ref{m-evolve}) and
Eq.~(\ref{f-evolve}) are the same we will only analyse the former
equation and apply results to the latter.

The first term in the square brackets indicates the contribution
to $A_{k}$ when a male of $k-1$ connections acquires  a new
contact, the corresponding losses are given by the second term.
The growth of the network due to influx is expressed by the last
term, with $p$ being a probability of addition of males with $k=1$
connection. The multiplicative factor $M_{a}$ is to ensure the
appropriate normalization and is given by \begin{eqnarray}
\label{M_{a}} M_{a}(t)&=&\sum_{k}(k+\lambda)A_{k}(t).
\end{eqnarray} The total number of males in the system is $A(t)=\sum_{k}A_{k}(t)$
and the total number of females $B(t)=\sum_{l}B_{l}(t)$. From
initial conditions we have $A(t)=1+pt$ and similarly $B(t)=1+qt$
making an assumption that there is one of each sex representatives
before the growth commences. In particular, the bipartite
structure \cite{graph-t,randgraphs} of the graph ensures that the
total number of connections in the system is the same for both
males and females, which means that $L(t)\equiv
K(t)=\sum_{k}kA_{k}(t)=t+1$, with one link before the first time
step. Also, the network does not contain monogamous individuals
after the first time step.

We start our analysis by looking at the moments of $A_{k}(t)$,
which are defined as
\begin{eqnarray} \label{Ak-moments}
M_{i}(t)&\equiv&\sum_{k}k^{i}A_{k}(t).
\end{eqnarray}
For the first few moments it is easy to show that
\begin{eqnarray}\label{first few moment}M_{0}(t)&=& M_{0}(0)+(q+r+p)t
\end{eqnarray} and
\begin{eqnarray}\label{first moment} M_{1}(t)=M_{1}(0)+pt.
\end{eqnarray} For large times the initial values of the moments become
irrelevant so that we get
\begin{eqnarray}\label{M_{a}norm} M_{a}(t)&=&[q+r+p(1+\lambda)]t,
\end{eqnarray}
which is a linear function of time and indicates that
Eq.~(\ref{m-evolve}) is also linear in time. Using this input we
can write $M_{a}(t)=m_{a}t$ implying $m_{a}=q+r+p(1+\lambda)$ and
$A_{k}(t)=a_{k}t$. Substituting these relations in
Eq.~(\ref{m-evolve}) leads to a time independent recurrence
relation
\begin{eqnarray} \label{a
recurrence} \left(\frac{m_{a}}{q+r}+k+\lambda\right)a_{k}=
(k-1+\lambda)a_{k-1} + \frac{m_{a}}{q+r}p\delta_{k1}.
\end{eqnarray}Iterating Eq.~(\ref{a recurrence}) we obtain
\begin{eqnarray} \label{a gamma}
a_{k}=\frac{\Gamma(k+\lambda)}{\Gamma(\frac{m_{a}}{q+r}+k+\lambda+1)}
\frac{\Gamma(\frac{m_{a}}{q+r}+\lambda+1)}{\Gamma(\lambda)}\frac{m_{a}}{q+r}p.
\end{eqnarray}
Using the former relation for the case of females, by simply
replacing the relevant parameters gives
\begin{eqnarray} \label{b gamma}
b_{l}=\frac{\Gamma(l+\mu)}{\Gamma(\frac{m_{b}}{p+r}+l+\mu+1)}
\frac{\Gamma(\frac{m_{b}}{p+r}+\mu+1)}{\Gamma(\mu)}\frac{m_{b}}{p+r}q.
\end{eqnarray}
For large $k$ and $l$ Eq.~(\ref{a gamma}) and Eq.~(\ref{b gamma})
are asymptotically equivalent to
\begin{eqnarray} \label{a asymptotic}
a_{k}\sim k^{-(1+\frac{m_{a}}{q+r})}
\end{eqnarray} and
\begin{eqnarray} \label{b asymptotic}
b_{l}\sim l^{-(1+\frac{m_{b}}{p+r})}.
\end{eqnarray}
The distributions $a_{k} \sim k^{-\gamma_{m}}$ and $b_{l} \sim
l^{-\gamma_{f}}$ scale as power-laws, which is a vulnerable
structure for social networks as far as the epidemics and the
spreading of computer viruses are concerned
\cite{virus,epidemics,aids}. The scaling exponent
\begin{eqnarray}\label{gama-m}\gamma_{m}=2+\frac{p(1+\lambda)}
{1-p}\end{eqnarray}depends upon male charisma parameter $\lambda$
and the rate $p$ of addition of males into the network. Similarly,
the exponent
\begin{eqnarray}\label{gama-f}\gamma_{f}=2+\frac{q(1+\mu)}{1-q}
\end{eqnarray} depends upon female charisma parameter $\mu$ as well as the rate of
arrival of females. However, since we have the constraint
$p+q+r=1$ implying one link at each time step, the growth of one
network is effected by the growth of the other.

%************************************************************
\section{Discussion and conclusions}
\label{sec:conclusions}

From the definition of the model, we have four free parameters,
$0\leq p\leq 1$, $0\leq q\leq 1$, $\mu>\lambda>0$ and the fifth
one is given by the relation $(p+q+r)=1$.

Starting with a simple example, let us ignore the process $r$ and
take $p+q=1$. This yields
\begin{eqnarray}\label{simple-example}\gamma_{f}=2+\frac{q(1+\mu)}{1-q}
\quad \hbox{and} \quad \gamma_{m}=2+\frac{(1-q)(1+\lambda)}{q}.
\end{eqnarray}
We get $\gamma_{f}\to 2$,  $\gamma_{m}\to \infty$ as $q \to 0$ and
$\gamma_{f}\to \infty$,  $\gamma_{m}\to 2$ as $q \to 1$. Although
this is an unrealistic assumption, it simply illustrates the
strong coupling of the influxes of males and females into the
system. A similar strong coupling is also observed in
\cite{dafang} where a more general coupled growing network is
considered.

If we pick $\lambda=2$, $\mu=3$ and substitute into the scaling
exponents in  Eq.~(\ref{gama-m}) and Eq.~(\ref{gama-f})
respectively we get $p\approx 0.30$ and $q\approx 0.28$ for the
corresponding power-law exponents in \cite{liljeros}. Where
$\gamma_{m}=1+\alpha_{m}=3.31$ and $\gamma_{f}=1+\alpha_{f}=3.54$
for the short time behaviour analysis. In this example choice of
values for $\lambda$  and $\mu$  allowed us to obtain $p \approx
q$ meaning that the increase of links due to new males and females
is approximately the same and about 40 \% of the links are created
among the pre-existing males and females. The average number of
partners of the sexes  $\sum_{k}kA_{k}/\sum_{k}A_{k}=1/p\approx 3
$ is the same since we have a close system.

Let us now consider the distribution of contacts during the life
time of the sexes. Here, we introduce another parameter $\beta$ to
the attachment rates to  describe bad experiences, aging or
inactivity of the sexes and hence the new parameter is a negative
input. It  can be in the range $[0,k+\lambda)$ or $[0,l+\mu)$,
which seems reasonable to assume for $k,l>2$. We can denote
$\lambda_{1}=\lambda-\beta_{m}$ and $\mu_{1}=\mu-\beta_{f}$ where
we assign a unique $\beta$ parameter to each kind to discriminate
their sexual behaviour. The new scaling exponents
\begin{eqnarray}\label{g-f}\gamma_{f}=2+\frac{q(1+\mu_{1})}{1-q}
\end{eqnarray} and similarly
\begin{eqnarray}\label{g-m}
\gamma_{m}=2+\frac{p(1+\lambda_{1})}{1-P}.
\end{eqnarray}
Using the same values we used earlier for the free parameters
gives $\gamma_{f}=1+\alpha_{f_{tot}}=3.1$ for $\beta_{f}\approx
1.17$ the same for all females and
$\gamma_{m}=1+\alpha_{m_{tot}}=2.6$ for $\beta_{m}\approx 1.6$ for
all males. Consequently we have $\beta_{f}<\beta_{m}$ implying
that the sexual connections of males tend to suffer more from
aging, bad experiences or inactivity than that of females. Perhaps
this may be the reason why males resort to lies to maintain self
image.

In conclusion, our simple model seems adequate to simulate the
dynamics of the web of human sexual contacts. The gratifying
feature of rate equation approach is that it gives flexibility to
include diverse range of parameters to model any situation within
the frame of this work. The charisma parameters used here are
shaped by the western social expectations and they can be changed
to reflect the views of other societies. Also, one can introduce
\emph{fitness}, multiplicative quenched disorder into the system
as in \cite{ergun} to account the activity level of different age
groups. Finally, a challenging issue to address would be to study
male-female degree correlations of this model.
%\************************************************************

\acknowledgments

I would like to thank EPSRC for financial support, Dafang Zheng,
Geoff Rodgers and  Yan Fyodorov for useful discussions.

%*************************************************************

\end{document}